\documentclass[twoside]{article}
\usepackage{pmfa0}
\usepackage{graphicx}
\usepackage[resetfonts]{cmap} 
\usepackage{lmodern} 
\usepackage[czech]{babel} 
\usepackage[utf8]{inputenc} 
\usepackage[T1]{fontenc} 
\usepackage{amsmath}

\begin{document}
\renewcommand\figurename{Obr.}

\title{Vliv sluneční aktivity na poruchy v české rozvodné síti: předběžné hodnocení}

\author{Tatiana Výbošťoková, Turčiansky Peter, Michal Švanda, Babice u Říčan}
\begin{poznamkapodcarou} {Bc. {\sc Tatiana Výbošťoková}, Astronomický ústav UK, MFF UK, V Holešovičkách 2, 180\,00 Praha 8, e-mail: {\tt
tatiana.vybostokova95@gmail.com},  doc. Mgr. {\sc Michal Švanda}, Ph.D., Astronomický ústav UK, MFF UK, V Holešovičkách 2, 180\,00 Praha 8, Astronomický ústav AV ČR, Fričova 298, 251\,65 Ondřejov, e-mail: {\tt michal@astronomie.cz}
}
\end{poznamkapodcarou}

\vspace{0.5cm}

Abstrakt: Eruptivní projevy sluneční aktivity ovlivňují bezprostřední kosmické okolí naší Země a postihují prostřednictvím indukce elektrických proudů i pozemskou infrastrukturu, zejména rozvodné sítě silové elektřiny. Teprve v~posledních letech byla věnována pozornost nejen hodnocení bezprostředních dopadů silných eruptivních událostí na zařízení rozvodné sítě, ale také statistickému zhodnocení vlivu zvýšené sluneční aktivity. Ve spolupráci s~ČEPS, a.s., jsme statisticky vyhodnotili četnost závad na zařízeních páteřní rozvodné sítě ČR v~závislosti na úrovni sluneční aktivity a ukazujeme, že vliv náhlých eruptivních událostí na závadovost zařízení české rozvodné sítě nelze vyloučit. 

\section{Eruptivní sluneční aktivita}
Slunce je nám nejbližší hvězdou, svými parametry (hmotnost $2\times 10^{30}$~kg, tedy 330\,000-krát více než Země, poloměr 696\,000~km, tedy 109-krát více než Země) je dominantním tělesem Sluneční soustavy. Jeho vliv však není omezen jen na gravitaci, jíž řídí pohyby všech těles v rámci planetárního systému, ale uvažovat musíme i vliv slunečního plazmatu, jímž Slunce vyplňuje okolní prostor. Koróna (vyšší vrstva atmosféry Slunce) je horká (s teplotou až 2 miliony kelvinů) a neustále se rozpíná do meziplanetárního prostoru ve formě tzv. slunečního větru. 

Sluneční vítr je složen především z nabitých částic -- iontů (dominují protony, dále pak jádra helia, tedy částice $\alpha$) i elektronů. Sluneční vítr má dvě základní komponenty, pomalou a rychlou, jež jsou doplněny občasnými výrony komponenty třetí, související s tzv. sluneční aktivitou. 

Slunce je objektem ve stavu plazmatu a je protkáno magnetickými poli. Tato pole jsou vytvářena a zesilována ve vnější obálce slunečního tělesa, vyvěrají z~něj a prolínají se sluneční atmosférou. Vzhledem k tomu, že svrchní obálka Slunce je velmi dynamická (především kvůli probíhající konvekci), podléhají i tato magnetická pole výrazným časovým změnám. Jevy, jež jsou spojeny s~existencí a proměnností lokalizovaných magnetických polí, označujeme souhrnným názvem \emph{sluneční aktivita}. Mezi ty nejznámější projevy patří jistě sluneční skvrny (chladnější a tedy temnější místa na slunečním povrchu) nebo protuberance (oblaka plazmatu vyplňující smyčky magnetických polí nad slunečním povrchem, jež mohou být dlouhodobě v rovnováze). Mezi nejdramatičtější projevy sluneční aktivity s potenciálním dopadem na pozemské prostředí patří erupce. 

Sluneční aktivita není v čase stálá, různé indexy aktivity (např. počet skvrn) vykazují semiperiodické chování s dominantními periodami v délce přibližně 11 let, 22 let a 86 let. První dvě zmíněné periody souvisejí s trendem aktivity kolísající z minima, kdy např. nejsou pozorovány žádné skvrny a i výskyt silných erupcí je ojedinělý, po maximum, v němž jsou na Slunci pozorovány denně desítky skvrn a kdy Slunce chrlí jednu erupci za druhou. V indexech aktivity však nalezneme i další harmonické komponenty i mnohé sekulární složky. 

\subsection{Erupce}
Sluneční erupce jsou spojeny s náhlým přepojením (rekonexí) magnetického pole \cite{Shibata}. Obvykle k ní dochází v~oblastech s lokálně zvýšenou aktivitou (často v oblastech se skvrnami), kde jsou všechny předpoklady pro existenci smyček magnetických polí vypínajících se vysoko do koróny. Při explozivní rekonexi dochází k prudkému ohřevu koronálního materiálu až na teploty desítek milionů stupňů, jež se stává zdrojem rentgenového záření. Vznikají urychlené svazky elektronů a protonů, které jednak bombardují hustší vrstvy sluneční atmosféry pod místem erupce, jednak jsou směrovány do meziplanetárního prostoru a stávají se jedním z faktorů ovlivňujících pozemské podmínky. A konečně se při erupci často uvolňuje odpojený oblak plazmatu (plazmoid) uzavřený ve svém vlastním magnetickém poli, jenž je rychlostí přesahující 1000 km/s vyvržen do meziplanetárního prostoru (mluvíme o tzv. výronu hmoty do koróny, angl. Coronal Mass Ejection, odtud známá zkratka CME). Meziplanetární CME mají největší potenciál silně ovlivňovat pozemské technologie.

\subsection{Geomagnetická aktivita}
Sluneční aktivita má prokazatelný vliv na stav meziplanetárního prostoru v okolí naší planety. Země není vystavena přímému vlivu nabitých částic, neboť je obklopena vlastní magnetosférou, jež tyto částice do značné míry stíní. Prostřednictvím tohoto pole však poruchy sluneční aktivity indukují poruchy v~magnetosféře, označované souhrnným názvem \emph{geomagnetická aktivita}, jejíž podstatné výkyvy označujeme jako geomagnetické bouře. 

Úroveň geomagnetické aktivity lze nejsnáze posoudit na základě měření intenzity zemského magnetického pole. Výkyvy sluneční aktivity lokálně mění magnetickou indukci o hodnoty v~řádu desítek až stovek nanotesla (změny jsou tedy sto až tisíckrát menší než je střední hodnota indukce), v klidovém stavu se indukce mění o méně než 20 nT. Změny větší již indikují situaci označovanou jako geomagnetická bouře. 

\subsection{Kvantifikace geomagnetické aktivity}
Geomagnetická aktivita je klasifikována prostřednictvím zavedených geomagnetických indexů. Tyto indexy vycházejí z aktuálních měření horizontální komponenty geomagnetického pole. Informaci o~aktuálním stavu podává DST index (Disturbance Storm Time index), jde o hodinový průměr odchylky horizontální složky měřený v blízkosti rovníku pozemními stanicemi. DST je mírou síly tzv. kruhového proudu kolem Země, jenž je způsoben protony a elektrony pocházejícími od Slunce. Záporné hodnoty DST značí zeslabení zemského magnetického pole. DST se používá ke klasifikaci probíhající geomagnetické bouře: je-li amplituda DST v rozsahu do 50 nT, jde o bouři slabou, jsou-li hodnoty v rozsahu 50 až 100 nT, jde o bouři střední, hodnoty 100--250 nT charakterizují bouři silnou a více než 250 nT pak označíme jako superbouři. 

V historii nejsilnější zaznamenanou geomagnetickou bouří je „Carringtonova superbouře“ ze září 1859, kdy DST index nabyl hodnoty $-1760$~nT (ta je však v~poslední době zpochybňována jako nadhodnocená \cite{Cliver}). Superbouře z května 1921 dosáhla DST $-850$~nT, v březnu 1989 pak $-589$~nT. Série bouří na konci října 2003 vykazovala hodnoty DST až $-383$~nT. 

Úroveň geomagnetické aktivity lze posoudit i s pomocí $K$-indexu. Jde o semilogaritmickou veličinu popisující změny amplitudy horizontální komponenty magnetického pole Země. Vypočte se z~hodnoty maximální fluktuace této komponenty v průběhu 3 hodin. Hodnota $K$-indexu 0 značí klidný stav, hodnoty 5 a vyšší geomagnetickou bouři, hraniční hodnota 9 značí superbouři. $K$-index charakterizuje poruchy v konkrétním místě.  Z celosvětové sítě geomagnetických observatoří je pak počítán vážený průměr vyhodnocující celoplanetární $Kp$ index. Geomagnetickou aktivitu ve vysokých šířkách pak charakterizuje spíše $AE$ index, související se změnami elektrického proudu v aurorálním oválu. 

\section{Ovlivnění rozvodných sítí}
Zvýšená geomagnetická aktivita má mnoho registrovatelných projevů, mezi něž patří např. výskyt polárních září -- výsledků kolizní excitace s následnou zářivou de-excitací atomů a molekul vzduchu. Výrazná aktivita polárních září je další známkou probíhající geomagnetické bouře a obvykle dobře koresponduje se zvýšenou hodnotou DST i $K$-indexu. Změny v ionizaci vyšších vrstev atmosféry mají vliv na průchod elektromagnetických vln, ovlivňováno je tak nejen šíření telekomunikačních signálů, ale i signály navigační, včetně rozšířeného GPS. Přímé expozici nabitými částicemi jsou vystaveny některé kosmické družice a zvýšené radiační dávky během geomagnetických bouří opakovaně obdrží i posádky letadel na dálkových trasách. 

Zemská magnetosféra i vyšší vrstvy atmosféry (ionosféra) obsahují systém elektrických proudů, jež podléhají v průběhu geomagnetické bouře značným změnám. Změny systému proudů generují časově proměnné elektrické pole, jež ve vodivých strukturách na povrchu Země budí parazitní elektrické proudy, tzv. geomagnetické indukované proudy (GIC).  GIC jsou buzeny nejen v přirozeném prostředí, ale také v~technologických prvcích, zejména v metalickém vedení telekomunikací, silovém vedení elektřiny nebo v~dálkových produktovodech. Pro představu: 25. 3. 1940 bylo v 48V telefonní síti společnosti AT\&T naměřeno indukované napětí 600 V, stejný den pak v transatlantických kabelech spojujících Evropu se Severní Amerikou až 2600 V. 

Za riziko s největším potenciálním dopadem je obecně považován kolaps rozvodné sítě (blackout). GIC v rozvodné síti vedou ke dvěma hlavním efektům:
\begin{enumerate}
\item GIC může být pojistnými prvky vyhodnocen jako nebezpečné přepětí, což vede k bezpečnostnímu odpojení postižené větve rozvodné sítě. To ovšem v praxi znamená, že celkový výkon přenášený sítí je následně směřován do méně větví, a to potenciálně včetně GIC. Může tedy dojít k zásahu pojistného prvku na jiné větvi atd. až do stavu, kdy není k dispozici dostatek zapojených vedení pro funkci sítě. Síť kaskádně kolabuje. Instalované filtrové banky často umožní regulovat vlastní GIC, ale již ne jeho harmonické frekvence. Kondenzátorové banky představují pro harmonické proudy cestu s nižší impedancí, na výsledný nadproud reaguje jistící prvek odpojením. 
\item GIC jsou sice proudy v čase proměnné, avšak jejich charakteristická perioda proměnnosti je několik málo minut, jsou tedy proměnné s frekvencemi v řádu mHz. Z hlediska zařízení typu transformátor, jejichž návrhová pracovní frekvence je 50~Hz, jsou GIC v podstatě proudy stejnosměrnými. Z toho vyplývá, že vnikne-li do transformátoru GIC, posunuje hysterezní křivku a saturuje jádro jednou polaritou. Transformátor se zahřívá, může plynovat (dochází k~termálnímu rozkladu olejové lázně na plyny jako vodík, metan nebo acetylén), poškozuje se mezidesková izolace, může dojít k požáru olejové lázně a v extrémních případech až k~tavení samotného jádra.
\item Přítomnost GIC ovlivňuje stabilitu frekvence systému, náhodné výkyvy představují zátěž pro turbíny generátorů a vznik nesouměrné soustavy fází kvůli přítomnosti zpětných proudů (zpětné soustavy). Systém ponechaný volnému vývoji může vést až k fyzickému poškození lopatek turbíny vznikajícími vibracemi. 
\end{enumerate}

\section{Světové události}
Na serveru {\it www.solarstorms.org} je zveřejněn archív novinových článků pojednávajících o zvýšené sluneční aktivitě. Pokud o těchto událostech přinášel informace i denní tisk, je zřejmé, že muselo jít o jevy skutečně významné. Archív obsahuje 306 článků o 105 událostech od roku 1859, přičemž při 60 z těchto událostí byly zaznamenány vlivy na pozemské technologie. 

1. září 1859 spatřili amatérští pozorovatelé Richard Carrington \cite{Carrington} a Richard Hodgson \cite{Hodgson} v oblasti slunečních skvrn bílý záblesk -- vůbec poprvé v historii pozorovali bílou erupci. Následující dva dny zasáhla Zemi silná geomagnetická bouře \cite{Cliver}. CME tedy vzdálenost Slunce--Země překonala za 17,6 hodiny. Vysoká cestovní rychlost byla zřejmě možná jen kvůli sérii předchozích CME, jež vyčistily meziplanetární prostor (o čemž svědčí i pozorování polárních září z~29.~srpna téhož roku). Silná geomagnetická bouře budila polární záře pozorovatelné ze subsaharské Afriky, Mexika nebo indické Bombaje. Ve Skalistých horách byly polární záře tak jasné, že horníci vstávali a připravovali si snídani v domnění, že je ráno a svítá. V Evropě a Spojených státech amerických zcela selhala telegrafní síť -- mnoho operátorů utrpělo od přetížených zařízení elektrické šoky, z telegrafních sloupů sršely jiskry až na zem.

15. května 1921 \cite{Silverman} vypadl celý signální a řídicí systém městské železnice v New Yorku, záhy následoval požár řídící věže na 57th Street a Park Avenue. Systém zkolaboval kvůli přítomnosti silných GIC, požár řídící věže byl zažehnut od telegrafu. Kabelová komunikace byla narušena ve většině Evropy, ve Švédsku byla hlášen požár telefonní ústředny.

22. ledna 1938 způsobila mohutná geomagnetická bouře problémy na železničním koridoru mezi Manchesterem a Sheffieldem, GIC pronikly do signálního zařízení a znemožnily jeho správnou funkci. 

13. března 1989 nastal blackout rozvodné sítě Hydroquébec v oblasti Hudsonova zálivu v Kanadě. GIC pronikly do páteřního 735kV vedení a síť kaskádně zkolabovala. Událost vedla k opatřením v síti Hydroquébec -- zvýšení úrovně přepětí nutného k~rozpojení stykače, instalaci kompenzátorů na síti velmi vysokého napětí a přepracování monitorovacích a operačních postupů. Kromě blackoutu byla zasažena důležitá zařízení, např. dva zvyšovací transformátory elektrárny La Grande 4 byly poškozeny přepětím v okamžiku odpojení zátěže v síti. Postiženy byly i další systémy v rozvodné síti nejen Kanady, ale i Spojených států a Evropy \cite{NERC}\cite{Kappenman}.

Zasaženy byly i dva bloky elektrárny Salem (New Jersey, USA) a jeden blok elektrárny Hope Creek. Průchodem GIC s odhadovanou vrcholovou amplitudou 224 A byla výrazně poškozena jádra zvyšovacích transformátorů na blocích Salem I a II, po demontáži byly zjištěny známky termální degradace izolace i~jasně viditelné tavení desek v jádře. 

Za zmínku jistě stojí, že v 25 měsících následujících po bouři z března 1989 havarovalo ve Spojených státech dvanáct tranformátorů.  

30. října 2003 vyvolaly GIC hodinu trvající blackout v oblasti Malmö ve Švédsku. Zaznamenané amplitudy GIC dosahovaly hodnot až 600 A. Ve dnech 29. až 30. října bylo v této oblasti zaznamenáno celkově deset výpadků různých zařízení sítě (odpojení přenosových tras, odpojení transformátorů) \cite{Lundstedt}.

Je zřejmé, že mohutné geomagnetické bouře mají bezprostřední vliv na stabilitu a funkčnost rozvodných sítí a bezchybnost instalovaných zařízení. Otázka, zda lze vysledovat vliv i slabších geomagnetických bouří, byla seriózně zodpovězena teprve v~nedávné době. Práce \cite{CM2013} a \cite{S2014} se zabývaly statistikou hlášenek poruch severoamerické rozvodné sítě i statistikou pojistných událostí souvisejících s fungováním rozvodné sítě. Obě práce nezávisle docházejí ke shodnému závěru, že 4~\% poruch v americké síti lze statisticky připsat sluneční aktivitě. Ukazuje se, že v 5~\% nejbouřlivějších dnů se počet hlášení pojistných události zvyšuje až o 20~\%, třetina nejbouřlivějších dní pak značí celkový nárůst pojistných hlášení o 10~\%. Extrapolací docházejí autoři k závěru, že ročně je 500 pojistných událostí souvisejících bezprostředně s provozováním americké rozvodné sítě vyvoláno vlivem sluneční aktivity. 

\section{Hodnocení vlivu sluneční aktivity na českou rozvodnou síť}
Přestože jsou známy poruchy zařízení rozvodných sítí spojených se silnými událostmi sluneční aktivity i v~Evropě, statistické vyhodnocení vlivu pro evropské země nikdo neprovedl. Důvodů může být několik, například to, že provést velkoplošnou studii pro srovnatelné území jako jsou Spojené státy je nesmírně obtížné vzhledem k~tomu, že každý evropský stát má svého vlastního operátora, který se se závadami vypořádává podle svých vlastních interních předpisů. Sestavení celkové přehledové statistiky by pak značilo využití mnoha nezávislých zdrojů s~nejasným napojením, což by práci mohlo komplikovat. My jsme provedli pilotní studii pro území České republiky. 

\subsection{Data}
\begin{figure}[!t]
\includegraphics[width=0.7\textwidth]{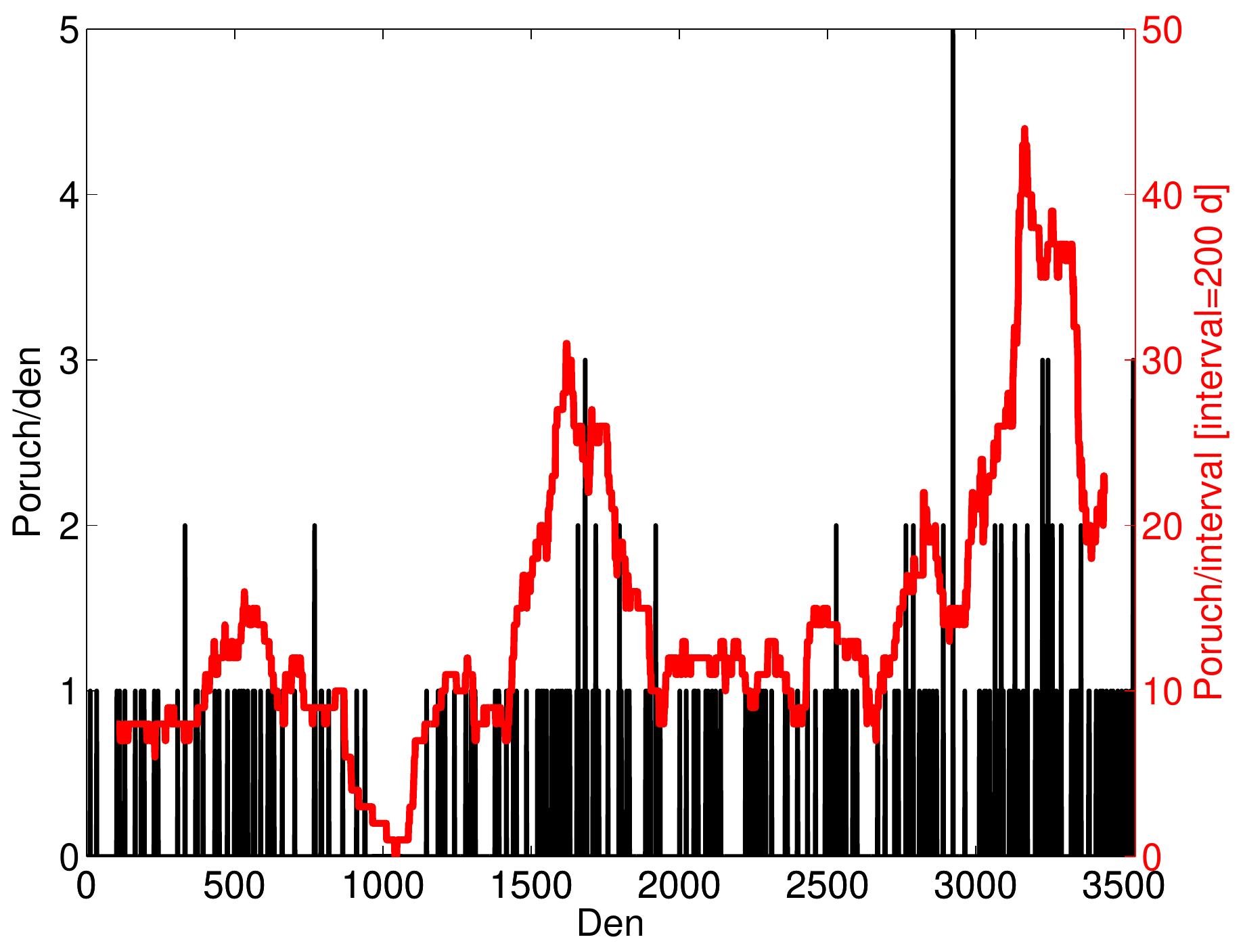}
\caption{Počet závad v~daném dni v~průběhu sledovaného období. Vzhledem k~charakteru křivky pro úplnost překreslujeme i zhlazenou verzi s~plovoucím oknem 200 dní, v~níž je dobře patrný dlouhodobý trend. Možnou souvislost tohoto dlouhodobého trendu s~celkovou úrovní sluneční aktivity jsme vzhledem k~celkové délce datové základny nevyšetřovali.}
\label{fig:CEPS}
\end{figure}

Společnost ČEPS, a.s., jež spravuje páteřní vedení české rozvodné sítě, nám poskytla seznam poruch na svých klíčových vedeních v~letech 2006--2015. V~tomto období zaznamenali technici celkově 338 závad nejrůznější závažnosti, z~nichž jsme pro další zpracování vyloučili ty, které prokazatelně nemohly být způsobeny externími vlivy. Mezi vyloučenými závadami byly např. závady, které nastaly před uvedením zařízení do provozu (tedy výrobní vady), nebo např. zatopení vodou. Do dalšího zpracování tedy vstupovalo 322 poruchových hlášenek. Byť jsou popisy závad velmi detailní, nás pro účely prvotního posouzení vlivu sluneční aktivity zajímalo pouze datum, kdy k~závadě došlo (viz obr.~\ref{fig:CEPS}). 

K~popisu úrovně geomagnetické aktivity jsme použili co možná nejbližší měřicí stanici, v~tomto případě geomagnetickou observatoř Budkov na Šumavě, provozovanou Geofyzikálním ústavem AV ČR, v.v.i. Prostřednictvím sítě InterMagnet jsou k~dispozici minutová měření plného vektoru geomagnetického pole. Z~těchto měření jsme zkonstruovali $K$-index, obvyklý pro charakterizaci úrovně geomagnetické aktivity v~podobných aplikacích. 

\begin{figure}[!t]
\includegraphics[width=0.7\textwidth]{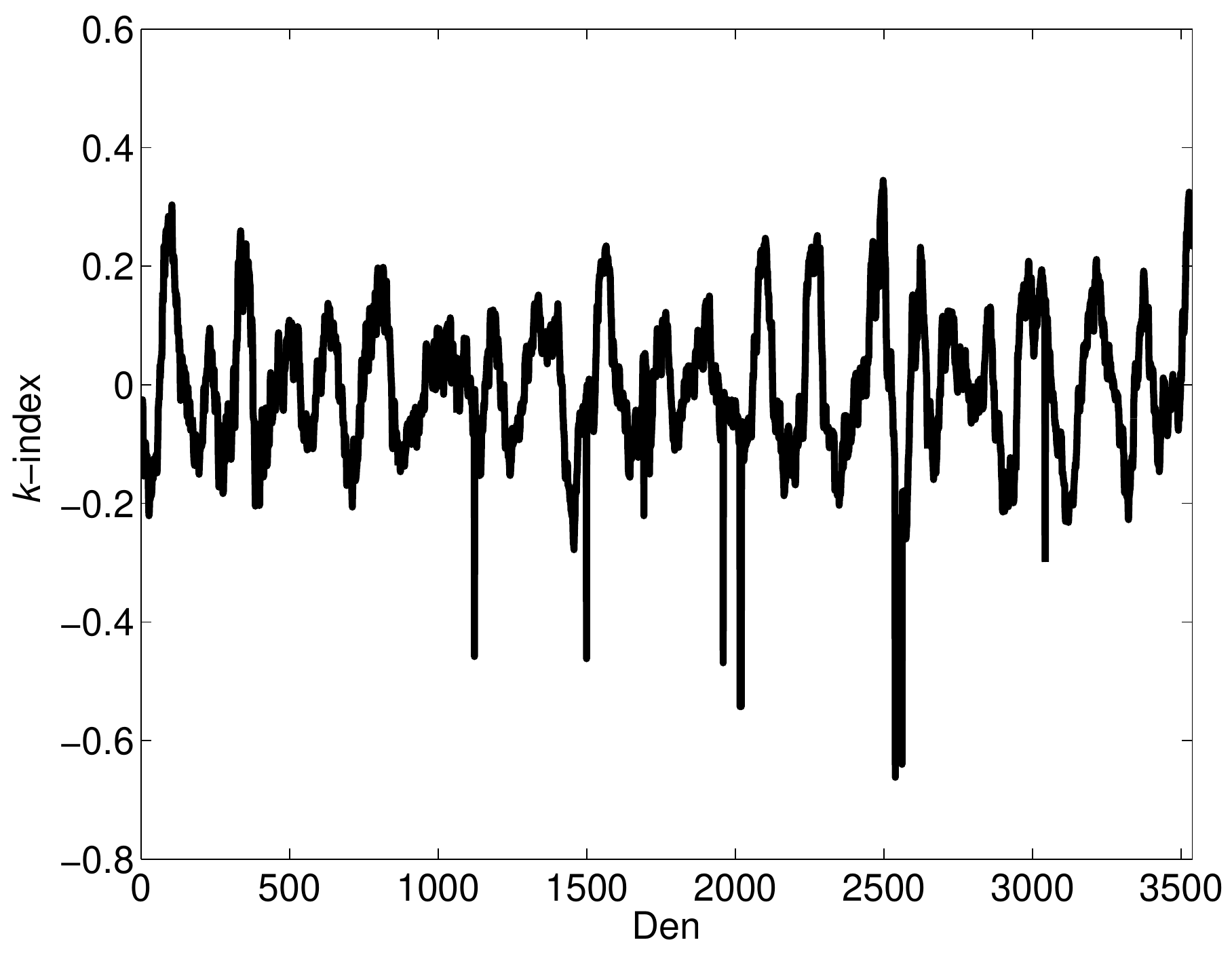}
\caption{Průběh filtrovaného $k$-indexu ve sledovaném období. }
\label{fig:k}
\end{figure}

Geomagnetické pole však výkyvy reaguje na mnoho vlivů, přičemž náhlé eruptivní události jsou jen jedním z~nich. Abychom odstranili pravidelné trendy související s~geometrií oběhu Země kolem Slunce a také se sluneční rotací, z~řady $K$-indexů jsme odstranili zřetelně přítomné periodicity o délkách jednoho roku a 28 dní. Tyto příspěvky nesouvisejí s~náhlými událostmi sluneční aktivity, o jejichž vliv jsme se zajímali především. Je nutné podotknout, že škála filtrovaného $K$-indexu (dále jen $k$-index), v~němž byla ponechána jen aperiodická složka, již fyzikálně neodpovídá původní škále významnosti geomagnetické bouře a může dokonce nabývat \clqq nefyzikálních\crqq{} záporných hodnot v~období extrémně snížené aktivity (viz obr.~\ref{fig:k}).  

\subsection{Metody vyhodnocování}
Cílem naší práce bylo porovnání obou datových řad. Korelační koeficient mezi $k$-indexem a křivkou počtu poruch, byť zhlazenou přes několik dní, je blízký nule -- tento výsledek jsme ostatně očekávali, jednak v~souladu se zjištěními v~podobných pracích ostatních autorů a jednak v~souladu s~fyzikálním náhledem, kdy případné poruchy nenastávají okamžitě po vstupu GIC do zařízení, ale s~nějakým časovým odstupem. 

Postupovali jsme tedy obezřetněji. Naším cílem bylo porovnat počet závad zaznamenaných ČEPS, a.s., ve třech různých typech časových intervalů: v~obdobích se zvýšenou geomagnetickou aktivitou, v~obdobích s výrazně sníženou geomagnetickou aktivitou a v~náhodně zvolených intervalech. Délky a počty těchto tří typů intervalů byly zvoleny stejné, abychom si zjednodušili jejich vzájemné porovnání statistickými metodami. 

Pokud platí pracovní hypotéza, že zvýšená geomagnetická aktivita se projeví v~počtu závad zaznamenaných na zařízení rozvodné sítě, pak by celkový počet závad $N_{\rm v}$ zaznamenaný v~obdobích zvýšené aktivity měl převyšovat počet závad $N_{\rm n}$ zaznamenaných v~období snížené aktivity. Dále bychom očekávali, že počet závad $N_{\rm r}$ v~náhodně zvolených intervalech bude mezi těmito dvěma extrémy, neboť náhodně zvolené intervaly mohou obsáhnout jak intervaly s~vyšší, tak s~nižší aktivitou. 

Je zřejmé, že výsledky mohou být ovlivněny délkou posuzovaných intervalů. Není naším cílem soustředit se na volbu optimální délky intervalu, proto jsme vyhodnotili výsledky pro sadu délek.

Pracovní hypotézu jsme testovali pomocí hypotézy alternativní a binomického testu. Alternativní hypotéza vyhodnocovaná testy tedy zní: Zvýšená geomagnetická aktivita se neprojeví na počtu zaznamenaných závad a zjištěné rozdíly jsou pouze dílem náhody. Porovnávali jsme tedy počet závad v~intervalech se zvýšenou a sníženou aktivitou, zvýšenou a náhodně zvolenými intervaly, a nakonec sníženou a náhodně zvolenými intervaly. 

Nechť např.~$N_{\rm v}$, $N_{\rm n}$ značí celkové počty závad v intervalech s vysokou, resp.~nízkou aktivitou. Celkový počet závad v obou obdobích je $n=N_{\rm v}+N_{\rm r}$. Pokud platí alternativní hypotéza, pak u každé závady je stejná pravděpodobnost ($p=1/2$), že se vyskytne v jednom ze dvou sledovaných období. Počty závad v obou obdobích jsou tedy binomické náhodné veličiny s parametry $n$ a $p$. Proto očekáváme, že $N_{\rm v}$ i $N_{\rm r}$ budou s velkou pravděpodobností blízké hodnotě $pn=n/2$. S ohledem na vztah $n=N_{\rm v}+N_{\rm r}$ stačí testovat výchylku pouze jedné z hodnot $N_{\rm v}$ a $N_{\rm r}$.  Použijeme oboustranný binomický test, neboť výchylka od očekávané hodnoty $pn$ může nastat na obě strany.

Pravděpodobnost, že náhodná veličina s binomickým rozdělením s parametry $n$ a $p$ nabývá hodnoty aspoň $x$, je rovna
\[\sum_{k=x}^n\binom{n}{k}p^k(1-p)^{n-k}.\]
Jelikož $p=1/2$, předchozí výraz zároveň udává i pravděpodobnost, že náhodná veličina nabývá hodnoty nejvýše $n-x$. Potom 
\begin{equation}
P(x)=2\sum_{k=x}^n\binom{n}{k}p^k(1-p)^{n-k}
\label{binom}
\end{equation}
je pravděpodobnost, že hodnota veličiny je aspoň $x$ nebo nejvýše $n-x$ a vzorec platí pro $x>n/2$. V našem příkladu za $x$ dosazujeme ${\rm max}(N_{\rm v}, N_{\rm n})$ a rovnice (\ref{binom}) tak vyjadřuje pravděpodobnost, s jakou jsou zjištěné rozdíly v počtu závad $N_{\rm v}$ a $N_{\rm n}$ dílem náhody. Hladinu statistické významnosti volíme jako 5~\%. To znamená, že alternativní hypotézu zamítneme, pokud $P(N_{\rm v})$ je menší než 0{,}05.


Hodnotu rizika zvýšených závad v~období zvýšené geomagnetické aktivity jsme poté vyhodnotili pomocí metody kontrolního vzorku (case-control study), jež je standardní metodou používanou např. při vyhodnocování účinnosti vakcín nebo při epidemiologických studiích. Jsou zvoleny dvě skupiny lišící se kauzálním atributem (v tomto případě expozice zvýšené geomagnetické aktivity) a je porovnáván počet pozitivních a negativních výsledků v~těchto dvou skupinách. V~našem případě jsme vyhodnocovali relativní riziko $R$ vypočtené jako
\begin{equation}
R={\frac{a}{a+b}}/{\frac{c}{c+d}},
\end{equation}
kde $a$ je počet dní se závadou, $b$ je počet dní bez závady, obojí pro intervaly se zvýšenou sluneční aktivitou. Pro intervaly se sníženou aktivitou je $c$ počet dní se závadou a $d$ počet dní bez závady. 
Relativní riziko nabývá hodnoty 1, pokud není mezi oběma skupinami lišícími se kauzálním atributem rozdíl. Je-li jeho hodnota menší než 1, znamená to, že častěji se pozitivní případy vyskytují ve skupině bez kauzálního atributu (tedy v rozporu s očekáváním), je-li větší než 1, vyskytují se pozitivní případy častěji ve skupině s kauzálním atributem. Hodnota 1{,}20 pak znamená, že v~případě pozitivní expozice kauzálnímu atributu (zvýšené aktivitě v~našem případě) je o 20~\% vyšší pravděpodobnost závady než v~případě kontrolním (tedy ve snížené aktivitě). 

\subsection{Výsledky}
Vytvořeným programem jsme v~průběhu $k$-indexu hledali vhodné intervaly délky $W$, jejichž střed se nacházel v~pozicích lokálních maxim $k$-indexu. Ke každému intervalu zvýšené aktivity jsme strojově dohledávali párový interval se sníženou aktivitou (se středem v lokálním minimum) o stejné délce $W$ tak, aby se oba intervaly nepřekrývaly a současně se nacházely co možná nejblíže. Touto volbou jsme eliminovali možný vliv např. rozvoje struktury sítě nebo jejích podstatných změn, stejně jako střídání ročních období na počet závad. Jako doplňující vzorek program opět automaticky vybral stejný počet zcela náhodně zvolených intervalů, opět s~délkou okna $W$. 

Pro všechny tři typy intervalů jsme spočetli odpovídající počet závad a vyhodnotili je statisticky metodami popsanými v~předchozí podkapitole. Výsledky jsou pro několik voleb délky intervalu $W$ shrnuty v~tabulce~\ref{tab:1}. 
\begin{table}[!ht]
\begin{tabular}{c|ccccccc}
\hline
$W$ [dnů] & $N_{\rm n}$ & $N_{\rm r}$ & $N_{\rm v}$ & $P_{\rm n-r}$ & $P_{\rm r-v}$ & $P_{\rm n-v}$ & $R$\\
\hline
10 & 11 & 6 & 19 & 0{,}33 & 0{,}01 & 0{,}20 & 1{,}73 \\
30 & 27 & 31 & 50 & 0{,}69 & 0{,}04 & 0{,}01 & 1{,}85\\
50 & 41 & 43 & 72 & 0{,}91 & 0{,}01 & 0{,}005 & 1{,}76\\
70 & 70 & 81 & 102 & 0{,}42 & 0{,}14 & 0{,}02 & 1{,}46\\
100 & 82 & 109 & 124 & 0{,}01 & 0{,}36 & 0{,}004 & 1{,}51\\
\hline
\end{tabular}
\caption{Statistická analýza souvislosti sluneční aktivity a počtu poruch na páteřní rozvodné síti ČEPS, a.s. Pro různě dlouhá okna $W$ jsou uvedeny sumární počty závad pro intervaly s~nízkou aktivitou ($N_{\rm n}$), vysokou aktivitou ($N_{\rm v}$) a v~náhodně vybraných intervalech ($N_{\rm r}$). Tabulka dále uvádí hodnoty pravděpodobnosti $P$, s~níž jsou zjištěné rozdíly dílem náhody, a relativní riziko $R$ závadovosti v~období zvýšené aktivity oproti období snížené aktivity. }
\label{tab:1}
\end{table}

Ze souhrnu vyplývá, že až na nejkratší testovaný interval s~délkou 10 dní pro všechny ostatní délky platí, že $N_{\rm n} < N_{\rm r} < N_{\rm v}$, což je v~souladu s~pracovní hypotézou. Statistickou významnost těchto rozdílů shrnují sloupce $P_{\rm n-r}$ vyhodnocující významnost rozdílu poruch v~intervalech s~nízkou aktivitou a náhodnými intervaly, $P_{r-v}$ mezi vysokou aktivitou a náhodnými intervaly a $P_{\rm n-v}$ mezi nízkou a vysokou aktivitou. Opět až na nejkratší interval 10 dnů je ve všech případech zjištěný větší počet závad v~období se zvýšenou geomagnetickou aktivitou ve srovnání s~obdobími s~nízkou aktivitou statisticky významný a nelze jej vysvětlit náhodnou realizací. S~výjimkou intervalů 10 a 70 dnů je statisticky významný i rozdíl mezi počtem závad registrovaných v~intervalech se zvýšenou aktivitou a v~náhodně volených intervalech.

Hodnota relativního rizika závad v~expozici zvýšené aktivitě je kolem 1,5 až 1,9, tedy v~období zvýšené aktivity je o polovinu až téměř o sto procent vyšší pravděpodobnost závady než v~období aktivity snížené. 

\subsection{Diskuse}
Z~naší předběžné studie vyplývá, že nelze vyloučit vliv zvýšené sluneční aktivity (vyhodnocované prostřednictvím indexu geomagnetické aktivity zbavené periodických trendů) na počty závad registrované v~české rozvodné síti, přinejmenším v~páteřní části spravované ČEPS, a.s. Z~tabulky~\ref{tab:1} se zdá, že nejvýhodnější délkou intervalu pro studium těchto jevů je 50 dnů -- to je ostatně v~souladu s~jedním z~výsledků studie \cite{CM2013}. Pro tento interval jsou rozdíly počtu závad v~intervalech s~expozicí a bez ní největší. 

I přesto nelze říci, že naše studie prokazuje vliv sluneční aktivity na českou rozvodnou síť, pouze poukazuje na souvislost, která však nemusí být příčinná. Pro posouzení přímé souvislosti je zapotřebí detailnější studie vyhodnocující závady nejen statisticky, ale posuzující je z~hlediska skutečných fyzikálních vlivů, např. vyhodnocením historie amplitud GIC v~daném místě. GIC sice nejsou v~zařízeních české rozvodné sítě monitorovány, ale je možné je matematicky modelovat podobně, jako to provedli např. Hejda a Bochníček \cite{Hejda} na ropovodu Družba během haloweenských bouří v~roce 2003. 

Možnou kauzalitu jsme předběžně posuzovali pomocí dodatečně provedeného testu. Pokud by zde byla patrná příčinná souvislost mezi zvýšením geomagnetické aktivity a zvýšenou závadovostí v~rozvodné síti, čekali bychom, že závady se budou kumulovat v~daném intervalu v~období po vrcholu aktivity, nikoli před. Testovali jsme tedy počet závad registrovaných ve stejně dlouhých intervalech délky $W$ kolem období zvýšené aktivity, přičemž tyto intervaly byly zvoleny tak, aby jeden interval končil v lokálním maximu $k$-indexu a druhý na něj v tomto maximu plynule navázal. Studované intervaly tedy neměly žádný překryv a jeden postihoval poruchy registrované těsně před lokálním maximem $k$-indexu a druhý těsně po tomto maximu. Pokud pracovní hypotéza kauzality platí, očekávali bychom, že v~intervalech pokrývajících období po maximu filtrovaného $k$-indexu budeme registrovat více závad než v~intervalu před lokálním maximem aktivity. 

\begin{table}[!ht]
\begin{tabular}{c|ccccccc}
\hline
$W$ [dnů] & $N_{\rm pred}$ & $N_{\rm po}$ & $P_{\rm po-pred}$ & $R$\\
\hline
10 & 12 & 19 & 0{,}28 & 1{,}58\\
30 & 43 & 50 & 0{,}53 & 1{,}41\\
50 & 59 & 72 & 0{,}29 & 1{,}22\\
70 & 79 & 102 & 0{,}10 & 1{,}29\\
100 & 110 & 124 & 0{,}40 & 1{,}13\\
\hline
\end{tabular}
\caption{Pro různé délky okna $W$ jsem vyhodnotili sumární počet závad těsně před a těsně po maximu $k$-indexu. Dále je uvedena pravděpodobnost $P$, že je tento rozdíl dílem náhody, a relativní riziko $R$. }
\label{tab:2}
\end{table}

Výsledky shrnuje tabulka~\ref{tab:2}. Z~ní vyplývá, že v~intervalech po lokálních maximech bylo skutečně registrováno více závad než před těmito maximy. Tyto rozdíly jsme opět vyhodnotili binomickým testem, který však ukazuje, že na hladině významnosti 5~\% jsou zjištěné rozdíly statisticky nevýznamné a nelze tedy zamítnout možnost, že jsou dílem náhody. Pro úplnost v~tabulce uvádíme i vypočtený relativní riziko z~metody kontrolního vzorku. 

Hlavním limitujícím faktorem naší pilotní studie je nízký počet zaznamenaných závad. V~budoucnu bychom tento faktor rádi vylepšili oslovením dalších operátorů a distributorů elektrické energie operujících na území ČR a snad též v~okolních státech. Vzájemná porovnávací studie by přinesla důležité indicie do nejisté kauzality mezi zvýšenou geomagnetickou aktivitou a zvýšenou závadovostí, neboť úroveň geomagnetické aktivity je v~regionu střední Evropy víceméně stejná. Jistě též nebude bez zajímavosti podívat se na závažnost závad v~exponovaných intervalech a porovnat je s~intervaly bez expozice. A konečně bychom rádi modelovali GIC v~exponovaných zařízeních a studovali jejich historii s~ohledem na vývoj stavu daného zařízení. 

\medskip

\noindent {\bf Poděkování}
Tento článek shrnuje výsledky bakalářské práce T. Výbošťokové vypracované pod vedením M. Švandy na Matematicko-fyzikální fakultě Univerzity Karlovy. Vysoce oceňujeme poskytnutí závadového deníku společností ČEPS, a.s., bez níž by tato studie nebyla možná, jmenovitě děkujeme zejména Petru Spurnému, DiS. Tato práce vznikla s podporou na dlouhodobý koncepční rozvoj vyzkumné organizace RVO:67985815.

{\small
\renewcommand \refname{\rm L\ i\ t\ e\ r\ a\ t\ u\ r\ a}

}

\end{document}